\documentclass[final]{IEEEtran}
\usepackage{amsthm,amssymb,graphicx,multirow,amsmath,color,amsfonts}
\usepackage[update,prepend]{epstopdf}
\usepackage[noadjust]{cite}
\usepackage[latin1]{inputenc}
\usepackage{balance}
\usepackage{tikz}
\usepackage{bbm} 
\usepackage{pdfpages}
\usepackage{tabulary}
\usepackage{multirow}
\usepackage{comment}
\usepackage{subfigure}

\def\nb0{{\mathbf{0}}}
\def\nb1{{\mathbf{1}}}

\newtheorem{lemma}{Lemma}

\newtheorem{theorem}{Theorem}

\allowdisplaybreaks 
\begin{document}
\title{
Stochastic Geometry-based Analysis of LEO Satellite Communication Systems}
\author{
Anna Talgat, Mustafa A. Kishk and Mohamed-Slim Alouini
\thanks{The authors are with King Abdullah University of Science and Technology (KAUST), Thuwal 23955-6900, Saudi Arabia (e-mail: anna.talgat@kaust.edu.sa; mustafa.kishk@kaust.edu.sa; slim.alouini@kaust.edu.sa)} 
}
\maketitle

\begin{abstract}
This letter studies the performance of a low-earth-orbit (LEO) satellite communication system where the locations of the LEO satellites are modeled as a binomial point process (BPP) on a spherical surface. In particular, we study the user coverage probability for a scenario where satellite gateways (GWs) are deployed on the ground to act as a relay between the users and the LEO satellites. We use tools from stochastic geometry to derive the coverage probability for the described setup assuming that LEO satellites are placed at $n$ different altitudes, given that the number of satellites at each altitude $a_k$ is $N_k$ where $1\leq k\leq n$. To resemble practical scenarios where satellite communication can play an important role in coverage enhancement, we compare the performance of the considered setup with a scenario where the users are solely covered by a fiber-connected base station (referred to as anchored base station or ABS in the rest of the paper) at a relatively far distance, which is a common challenge in rural and remote areas. Using numerical results, we show the performance gain, in terms of coverage probability, at rural and remote areas when LEO satellite communication systems are adopted. Finally, we draw multiple system-level insights regarding the density of GWs required to outperform the ABS, as well as the number of LEO satellites and their altitudes. 
\end{abstract}
\begin{IEEEkeywords}
Stochastic geometry, binomial point process, distance distribution, coverage probability.
\end{IEEEkeywords}

\section{Introduction} \label{sec:intro}
Satellite communications have a great potential to achieve the ultimate goal of providing wireless coverage worldwide, including rural and remote areas which are still lacking proper service around the globe~\cite{doi:10.1002/sat.941,9042251,9072129,dang_6g}. Particular attention has been recently given to LEO satellites for providing wireless coverage, due to their relatively low latency and cheaper launching costs~\cite{ref1}. This has motivated various companies to invest in launching large number of LEO satellites with the purpose of providing satellite-based cellular service, such as SpaceX, Amazon, and OneWeb~\cite{ref2}.

In this letter, using tools from stochastic geometry, we study the performance of LEO satellite-based communication systems given the altitudes of the satellites, the number of satellites at each altitude, and the density of the GWs. For that setup, we derive the coverage probability as a function of the aforementioned system parameters. In addition, we consider a practical use case where satellite communication systems are deployed to enhance coverage in remote and rural areas. For that purpose, we compare the coverage probability of the satellite-based communication system in such regions with the coverage probability in case of relying on the nearest ABS, which is typically located at far distances from rural and remote areas. Before providing more details regarding the contributions of this paper, we first discuss the related literature in the next subsection.

\subsection{Related work}
Performance analysis of satellite communication systems is essential for efficient implementation and design of such systems. Various aspects of such systems are investigated in literature such as the influence of the elevation angle and the altitude on the coverage area~\cite{ref4}, the influence of adopting non-orthogonal multiple access (NOMA)~\cite{ref6}, secrecy in UAV-aided satellite-terrestrial systems~\cite{8972910}, and overlay spectrum sharing~\cite{7981353}. Furthermore, many works have focused on deriving an accurate channel model for the communication links between satellites and ground stations~\cite{ref3,ref7}, where It was shown that shadowed-Rician (SR) model is the most accurate. One of the closest works in literature is~\cite{9079921}. In this work, authors studied a setup where the locations of the LEO satellites are modeled as a BPP on a spherical surface. Compared to the work in~\cite{9079921}, this paper has the following differences: (i) we consider a more general setup where satellites are deployed at various altitudes (multiple concentric spherical surfaces), which enables more general results, (ii) we adopt SR fading for the channel between the satellite and the GW, instead of Rayleigh fading adopted in~\cite{9079921}, and (iii) we provide useful insights for a specific use case of satellite communications in rural and remote areas. More details about the contributions of this paper are provided next.
\subsection{Contributions}
\begin{table*}
\centering
\caption{Table of Notations}
\label{tab:example}
\centering
\resizebox{0.85\textwidth}{!}{
\renewcommand{\arraystretch}{1.2}
\begin{tabular}{c|c}
    \hline
    \textbf{Notation } &  \textbf{Description}\\
    \hline
    \hline
    $N_i$ &  number of satellites on $i^{th}$ sphere   \\
    \hline
       $\sigma_g^2$; $\sigma_u^2$ &  Noise power at the GW; at the user   \\
    \hline
       $\gamma_g$; $\gamma_u$ &  SNR threshold at the GW; at the user   \\
    \hline
           $\rho_s$; $\rho_g$; $\rho_a$ &  Transmit power of the satellite; the GW; the ABS   \\
    \hline
    $r_e$; $r_i$; $a_i$ & Raduis of the Earth; radius of $i^{th}$ sphere; altitude of $i^{th}$ to the surface of the Earth.  \\
\hline 
    $D$; $R_{\rm GW-U}$; $R$ & Length of S-GW link; length of GW-U link; length of ABS-U link\\  
   \hline
    $P_{cov}^{\rm S-GW}$; $P_{cov}^{\rm GW-U}$; $P_{cov}^{\rm ABS-U}$;   & Coverage probability for  S-GW link; GW-U link; ABS-U link  \\
    \hline
    $W_s^2$; $W_g^2$; $D$ & SR fading power for the S-GW link; Rayleigh fading power for the GW-U link; Distance between GW and nearest Satellite\\
    \hline 
    $\mathcal{S}\mathcal{R}\left(\Omega, b_{0}, m\right)$ & $\Omega$ is 
    The line-of-sight component; $2 b_{0}$ is 
    the scatter component; $m$ is the Nakagami parameter.\\
    \hline
    \hline  
 \end{tabular}}
\end{table*}
We consider a system setup where the locations of the satellite GWs are modeled as a Poisson point process (PPP) on the ground and the LEO satellites are modeled as BPP on a set of spherical surfaces. In particular, to resemble practical scenarios, we assume that satellites are deployed on $n$ spherical surfaces $S_k$ where the number of satellites on $S_k$ is $N_k$ for $1\leq k \leq n$. For that setup, we study the coverage probability of a typical user on the ground as the joint coverage probability of the user-GW link and the GW-satellite link. Furthermore, we compare the coverage of such setup with the scenario where coverage is provided by a terrestrial base station (referred to as anchored base station or ABS in the rest of the paper) that is located far from the user, which resembles typical challenges in rural and remote areas. Using numerical results, we show the deployment of GWs to enable satellite communications enhances coverage probability when the distance to the nearest ABS is beyond a specific threshold. Finally, we show how this distance threshold is affected by various system parameters such as the density of the GWs, the altitudes of the satellites, and their numbers.
\section{System Model} \label{sec:SysMod}
\begin{figure}
\centering
\includegraphics[width=7 cm, page=1]{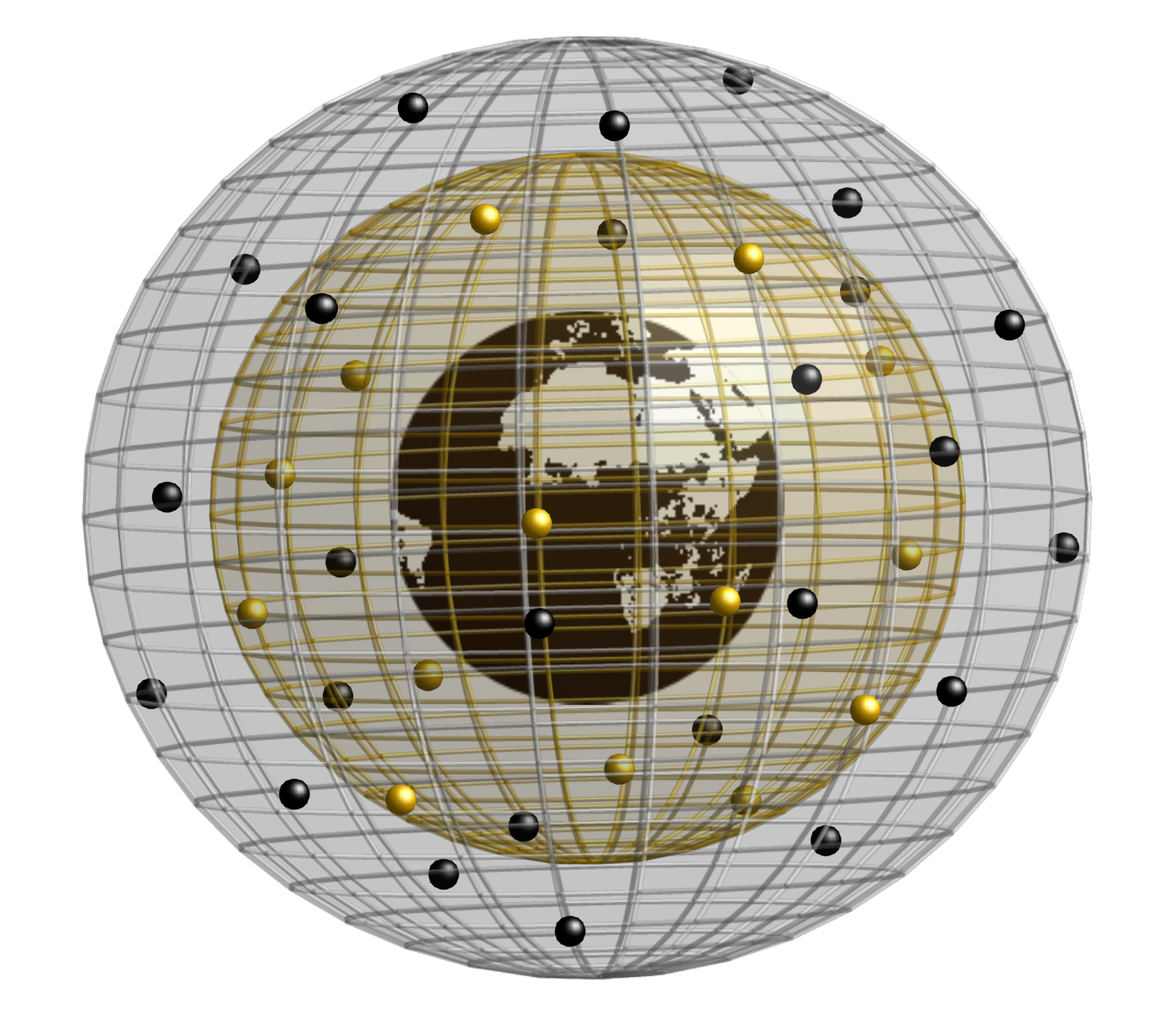}
  \caption{System model for $n$ level of spheres concentric with the Earth }
  \label{fig:1}
\end{figure}
We consider a setup where LEO satellites are deployed at $n$ different altitudes. Each altitude $a_k$ constitutes a spherical surface $S_k$ over which $N_k$ LEO satellites are uniformly distributed. The radius of each sphere $S_k$ is $r_k=r_e+a_k$, where $r_e$ is the radius of the earth. The communication between the LEO satellites and the ground users is relayed through GWs located on the ground. The locations of the GWs are modeled as a PPP with density $\lambda_{\rm GW}$. The transmit power of the satellite is denoted by $\rho_s$ while the transmit power of the GW is $\rho_g$. In the rest of the paper, we will refer to the link between the satellite and GW as the S-GW link while the link between the GW and the user will be referred to as the GW-U link. 
\subsection{S-GW Link}
For the S-GW link, the received signal power at the GW is
\begin{align}
\label{eq:power_S_G}
\rho_{r}^{g}=\rho_s|H_s|^2,
\end{align}
where $H_s$ represents the channel fading for the S-GW link and is represented as
\begin{align}
H_s=A\times W_s,
\end{align}
where $A$ and $W_s$ represent the propagation loss and the SR fading, respectively. The propagation loss for the S-GW link is computed using the below expression
\begin{align}
A=\frac{\lambda G_R  s e^{j \phi}\xi }{4\pi D},
\end{align}
where  $\lambda$ denotes the carrier wavelength, $D$ is the distance between the GW and the nearest satellite, $G_R^2$ represents the GW receiver antenna gain, $s$ denotes the rain attenuation coefficient, $\phi$ represents the phase due to the beam radiation pattern and radio propagation, and $\xi$ is a function of the maximum satellite antenna gain and the antenna bandwidth. 

The cumulative distribution function (CDF) of the SR fading power $W_s^2$ is given as follows:
\begin{align}
F_{W_{s}^{2}}(t)=\left(\frac{2 b_{0} m}{2 b_{0} m+\Omega}\right)^{m}\sum_{z=0}^{\infty} \frac{(m)_{z}}{z !\Gamma(z+1)} &\left(\frac{\Omega}{2 b_{0} m+\Omega}\right)^{z}\\ \nonumber
& \times \gamma\left(z+1, \frac{1}{2 b_{0}} t\right),
\end{align}
where $\Gamma(\cdot)$ denotes the gamma function, $\gamma(\cdot, \cdot)$ is the lower incomplete gamma function, $(m)_{z}$ is the Pochhammer symbol, while $m$, $b_0$ and $\Omega$ are the parameters of the SR fading.
\subsection{GW-U Link}
For the GW-U link, the received signal power at the user is 
\begin{align}
\label{eq:power_G_U}
\rho_r^{u}=\rho_g W_g^2 R_{\rm GW-U}^{-\alpha},
\end{align}
where $W_g^2$ is exponentially distributed with unity mean, $R_{\rm GW-U}$ is the distance between the GW and the user, and $\alpha$ is the path-loss exponent.
\subsection{Association Policy}
We consider an association policy where the user associates with its nearest GW, while the GW associates with its nearest LEO satellite. Hence, the user is considered in coverage if the following conditions are satisfied: (i) the signal-to-noise-ratio (SNR) for the S-GW link is above a predefined threshold $\gamma_g$ and (ii) the SNR for the GW-U link is above a predefined threshold $\gamma_u$. Hence, the coverage probability is defined as follows
\begin{align}\label{eq:cov}
P_{\rm cov}=P_{\rm cov}^{S-GW}P_{\rm cov}^{GW-U},
\end{align}
where 
\begin{align}
P_{\rm cov}^{S-GW}&=\mathbb{P}\left(\frac{\rho_r^g}{\sigma_g^2}\geq\gamma_g\right),\\
P_{\rm cov}^{GW-U}&=\mathbb{P}\left(\frac{\rho_r^u}{\sigma_u^2}\geq\gamma_u\right),
\end{align}
$\sigma_g^2$, and $\sigma_u^2$ are the noise powers at the GW and the user, respectively.

\section{Coverage Probability}
\subsection{Distance Distribution}
Given that the satellites are randomly located on the set of spheres $\{S_k\}$, the distance $D$ is a random variable. In the authors' work~\cite{talgat2020}, the distribution of $D$ was derived as shown in the below lemma.
\begin{lemma}[ Contact distance distribution] \label{lemma:1} 
\begin{align}
  F_{D}(d)  \overset{\Delta}{=} P(D< d)= 1- \prod_{i=1}^{n} P(D_i \geq d),
\end{align}
where the complementary CDF (CCDF) of $D_i$ is\\
$P( D_i\geq d)=$
\[\left\{\begin{array}{ll}
1,&d<a_i\\
\big[1-\frac{1}{\pi}\arccos{(1-\frac{{d^2-a_i^2}}{2r_er_i})})\big]^{N_i}, & a_i\leq d\leq d_{\rm max}(i,0)\\
\big[{1-\frac{1}{\pi}\arccos(\frac{r_e}{r_i})}\big]^{N_i},& d>d_{\rm max}(i,0),
\end{array}\right.\]
where $d_{\rm max}(i,0)=\sqrt{2r_e a_i+a_{i}^2}$. The PDF of $D$ is
\begin{align}
f_{D}(d)&=\left[\prod_{i=1}^{n}\Bar{F}_{D_i}(d)\right] \left[\sum_{i=1}^{n}\frac{f_{D_i}(d)}{\Bar{F}_{D_i}(d)}\right],
\end{align}
where 
\begin{align*}
f_{D_i}(d)&= \frac{d N_i}{\pi r_e r_i}[1-\frac{1}{\pi}\arccos({1-\frac{d^2-a_i^2}{2 r_i r_e}})]^{N_{i}-1} \\
&\times \frac{1}{\sqrt{1-(1-\frac{d^2-a_i^2}{2 r_e r_i})^2}},
\end{align*}
for $a_i\leq d\leq d_{\rm max}(i,0)$ and $f_{D_i}(d)=0$ otherwise.
\end{lemma}
\subsection{Coverage Analysis}
As described in (\ref{eq:cov}), in order the to compute the coverage probability, it is required to derive each of $P_{\rm cov}^{S-GW}$ and $P_{\rm cov}^{GW-U}$. Firstly, $P_{\rm cov}^{GW-U}$ is a well-established result in literature and is provided in the following lemma for completeness.
\begin{lemma}[] \label{lemma:2} 
The coverage probability for the GW-U link is
\begin{align}
P_{cov}^{\rm GW-U}=\int_{0}^{\infty}{\exp{\left(-\frac{\gamma_{u}r^{\alpha}\sigma_u^2}{\rho_g}\right)} f_{R_{\rm GW-U}}(r)}dr,
\end{align}
 where $f_{R_{\rm GW-U}}(r)=2\pi \lambda_{\rm GW}r \exp{(-\pi \lambda_{\rm GW}r)} $.
\end{lemma} 
Now, the main result in this paper, which is the derivation of $P_{\rm cov}^{S-GW}$, is provided in the below theorem.
\begin{theorem}[]\label{thm:2} 
The coverage probability for the S-GW link is
 \begin{align}
P_{cov}^{\rm S-GW}&=\int_{0}^{\infty}\frac{1}{2 \sqrt{y}}f_{D}(\sqrt{y})\\ \nonumber
&-\left(\frac{2 b_{0} m}{2 b_{0}m+\Omega}\right)^{m} \sum_{z=0}^{\infty} \frac{(m)_{z}}{z!\Gamma(z+1)}\left(\frac{\Omega}{2 b_{0}m+\Omega}\right)^{z}\\ \nonumber &\times \int_{0}^{\infty} \gamma\left(z+1, \frac{1}{2 b_{0}} cy\right)\frac{1}{2 \sqrt{y}}f_{D}(\sqrt{y})dy,
 \end{align}
 where $f_{D}(\sqrt{y})$ is given in Lemma~\ref{lemma:1} and $c=\frac{16 \pi^2 \gamma_{g} \sigma_g^2}{\rho_s| \lambda G_R s e^{j \phi}  \xi |^2}$.
 \begin{IEEEproof}
See Appendix~\ref{app:1}.
\end{IEEEproof}
\end{theorem}
\subsection{Coverage Enhancement in Remote locations}
\begin{figure}
\centering
\includegraphics[width=\columnwidth,page=2]{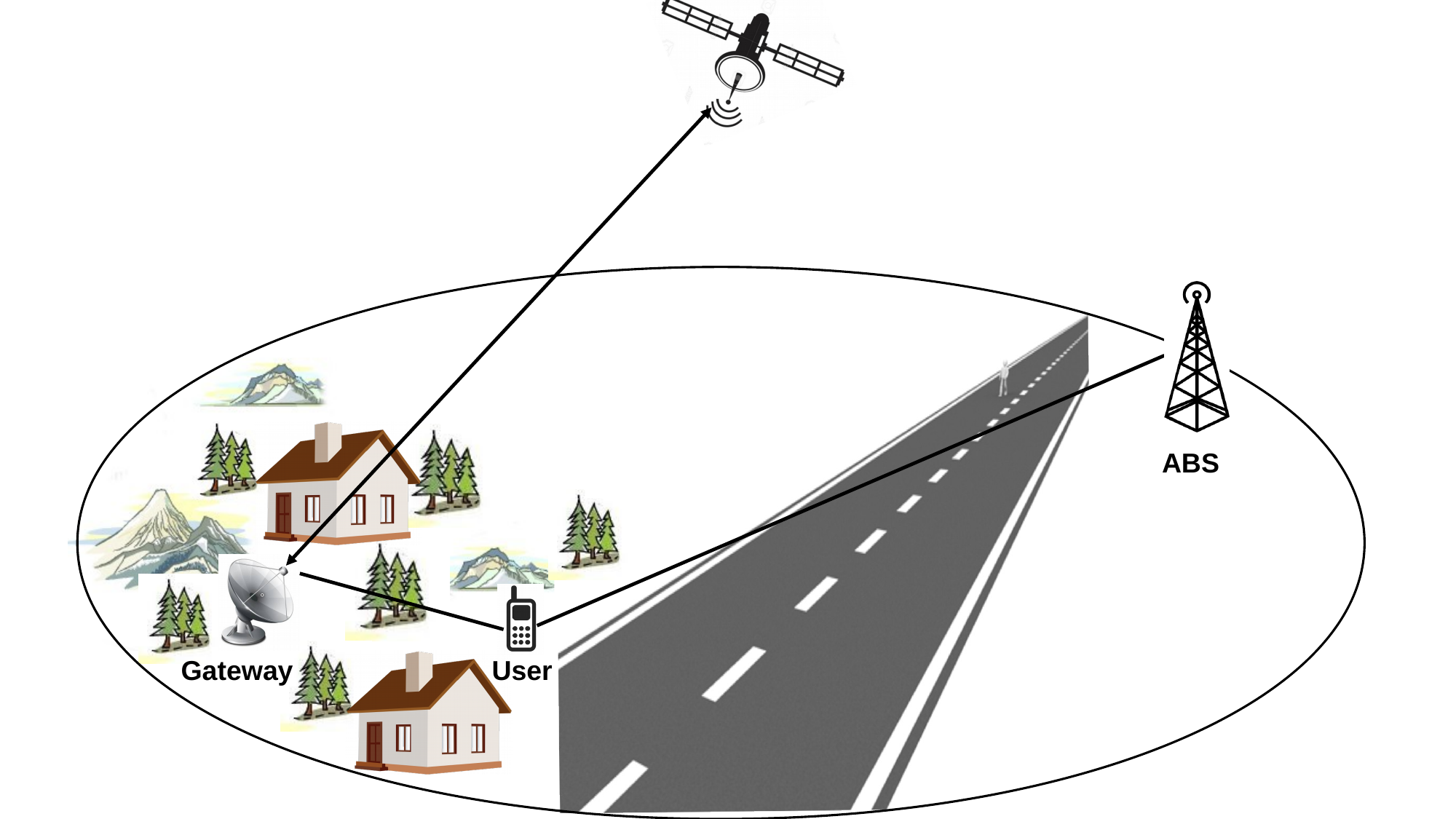}
\caption{ Satellite communication systems can highly enhance coverage for under-served remote and rural areas. }
\label{fig:2}
\end{figure}
In order to evaluate the performance gains from the deployment of satellite communication systems in rural and remote areas, we consider a scenario where, for a given rural area, the nearest ABS is located at a distance $R$, as shown in Fig.~\ref{fig:2}. The value of $R$ is typically large at rural and remote areas, leading to relatively low coverage probability. In such scenario, the coverage probability when the satellite communication system is absent is
\begin{align}
P_{cov}^{\rm ABS-U}=\exp{\left(-\frac{\gamma_{u}R^{\alpha}\sigma_u^2}{\rho_a}\right)} ,
 \end{align}
 where $\rho_a$ is the transmission power of the ABS and assuming Rayleigh fading. The above scenario will be used as a benchmark to evaluate the performance enhancement in the numerical results section, which is provided next.
\section{Numerical Results} 
\begin{figure*}[!t]
\centering
\subfigure[]{\includegraphics[width=0.3\textwidth]{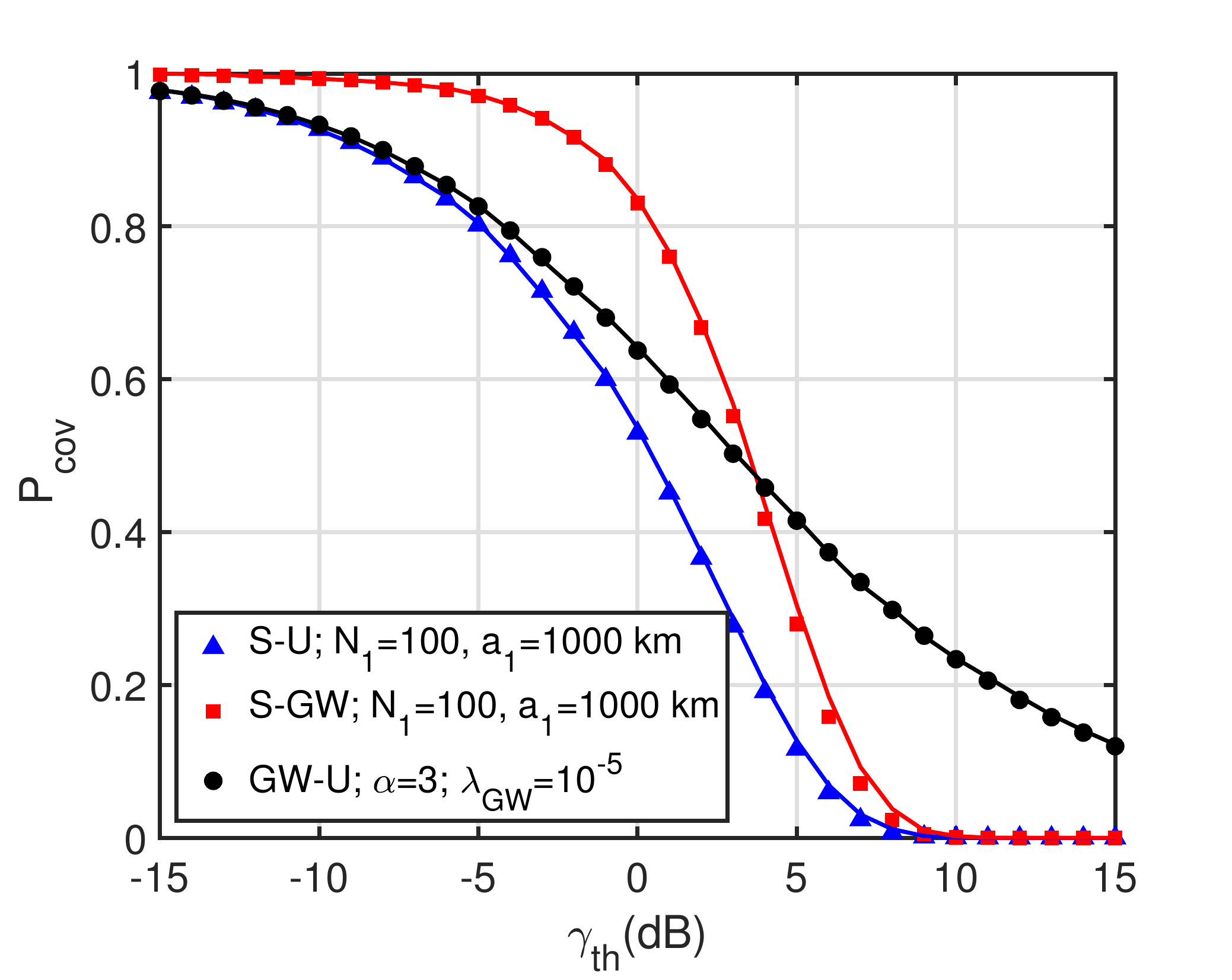}\label{3a}}
\subfigure[]{\includegraphics[width=0.3\textwidth]{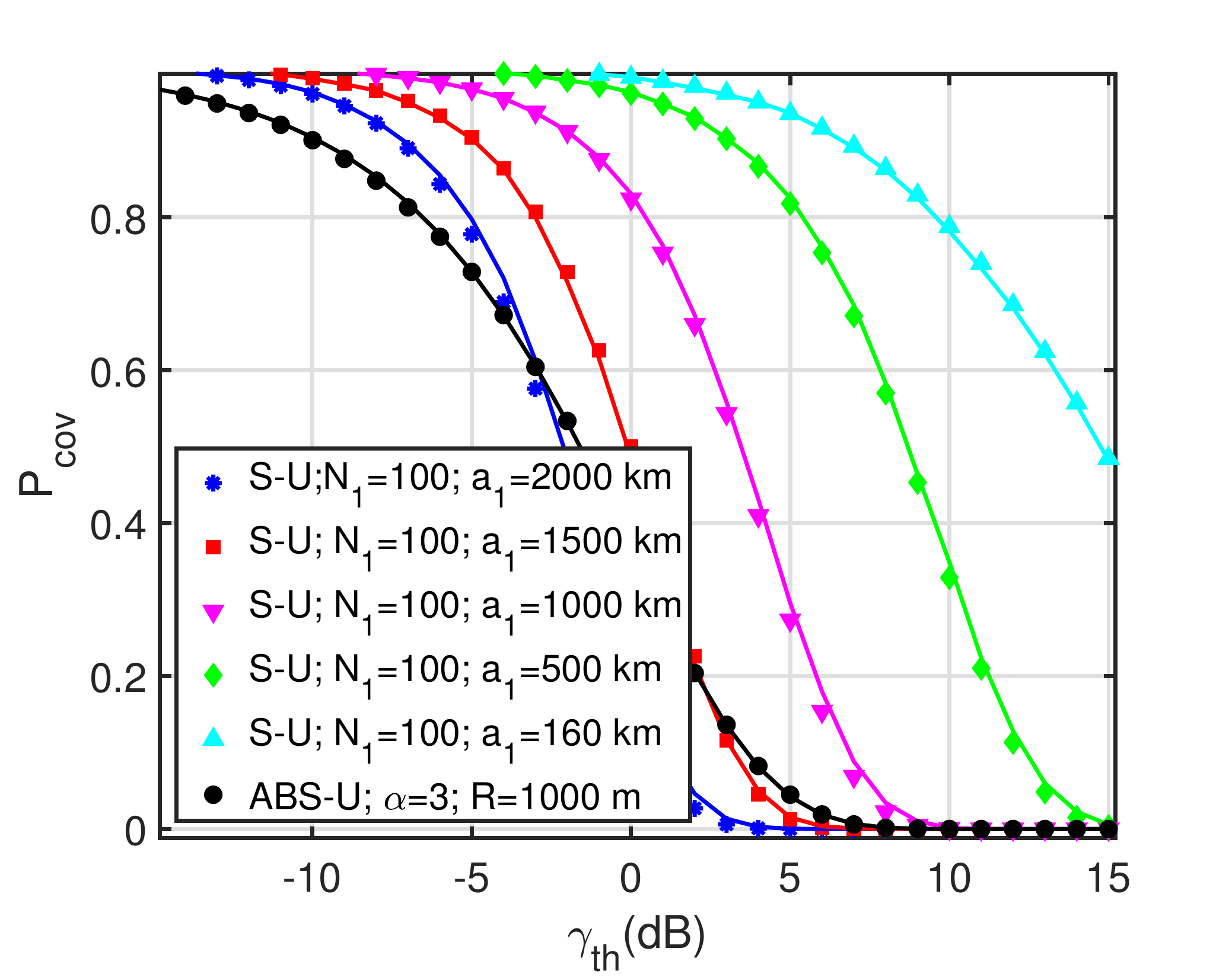}\label{3b}}
\subfigure[]{\includegraphics[width=0.3\textwidth]{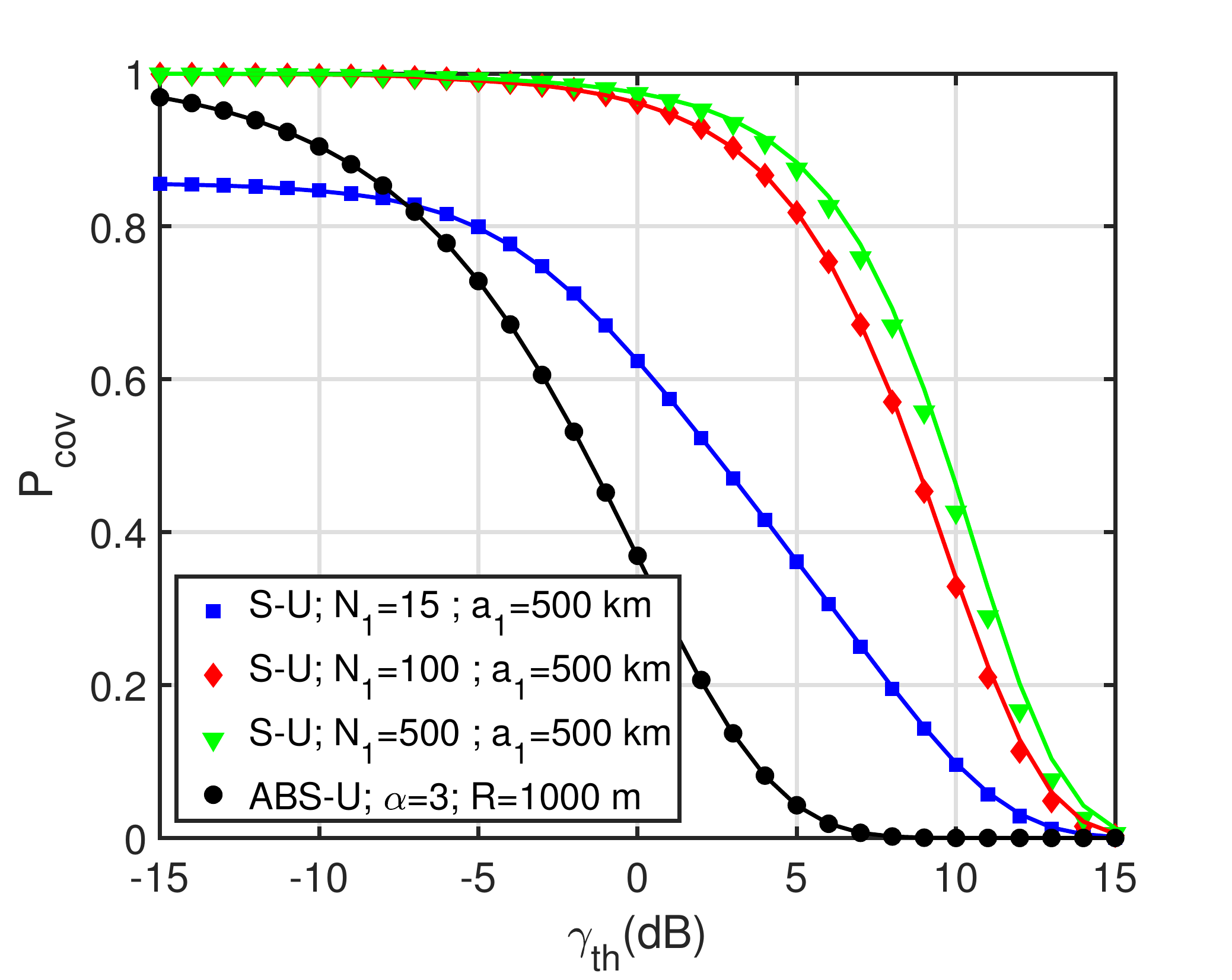}\label{3c}}
\caption{Coverage probability versus threshold, $\gamma_{th}$}
\label{fig:3}
\end{figure*}
\begin{figure*}[!t]
\centering
\subfigure[]{\includegraphics[width=0.3\textwidth]{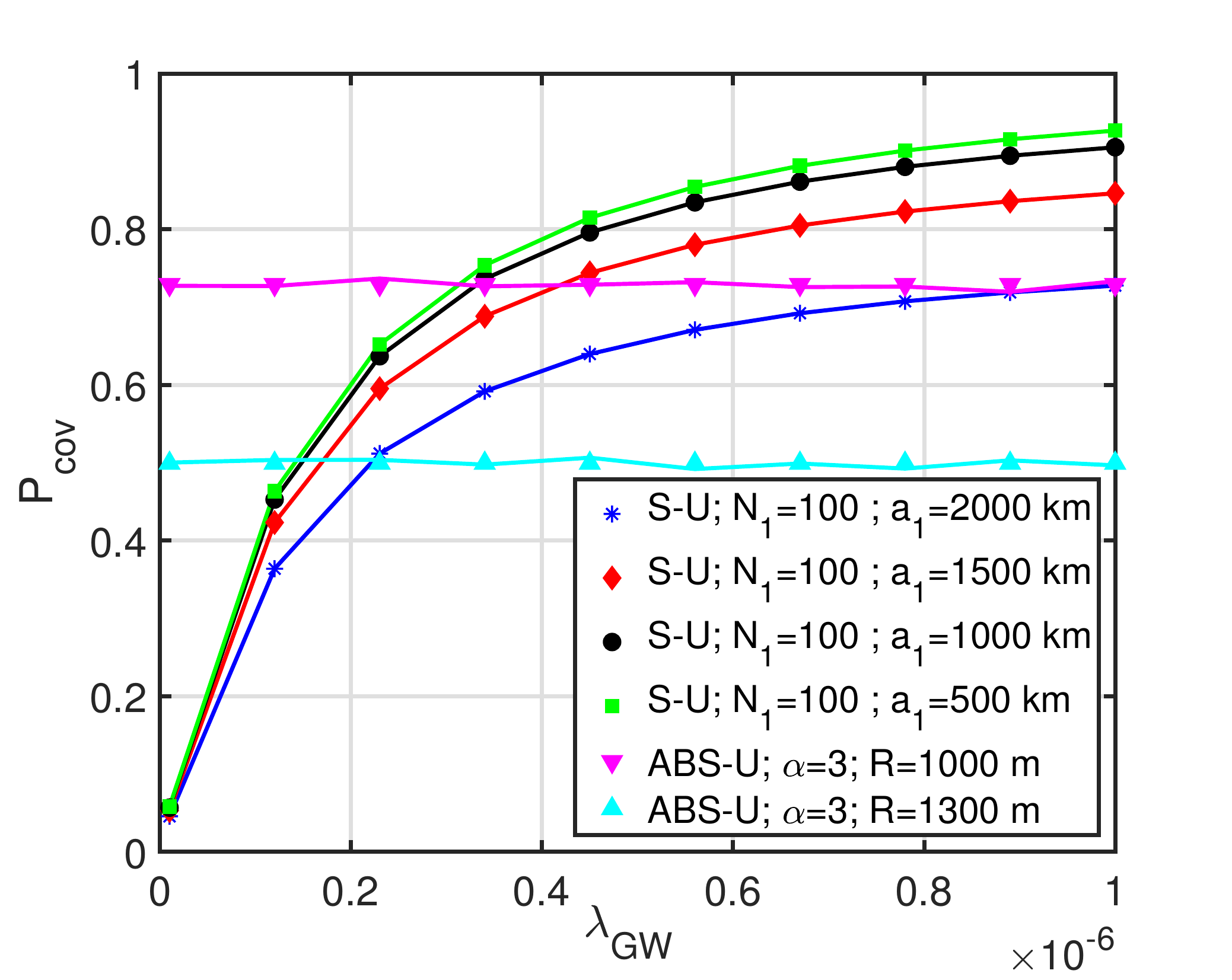}\label{4a}}
\subfigure[]{\includegraphics[width=0.3\textwidth]{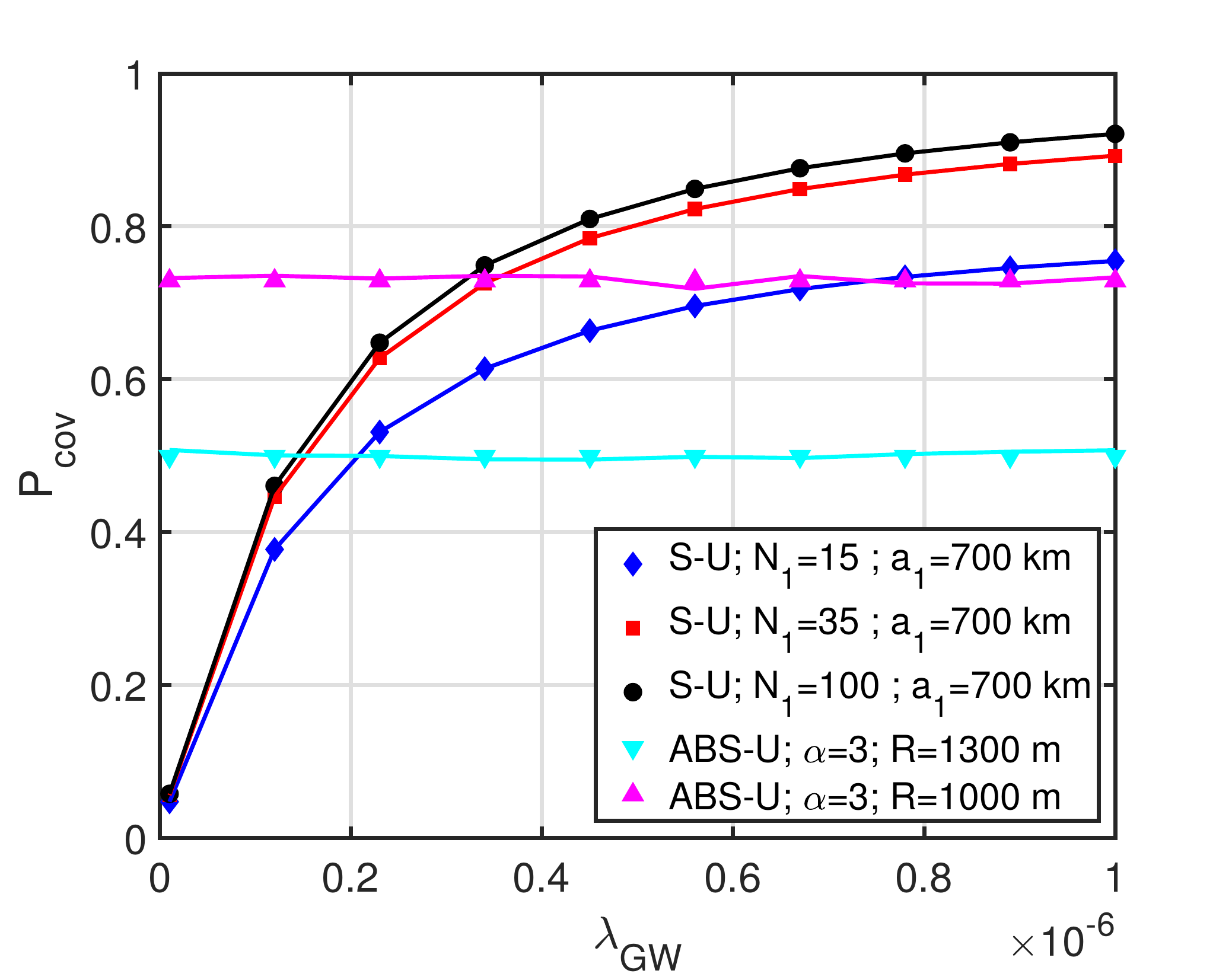}\label{4b}}
\subfigure[]{\includegraphics[width=0.3\textwidth]{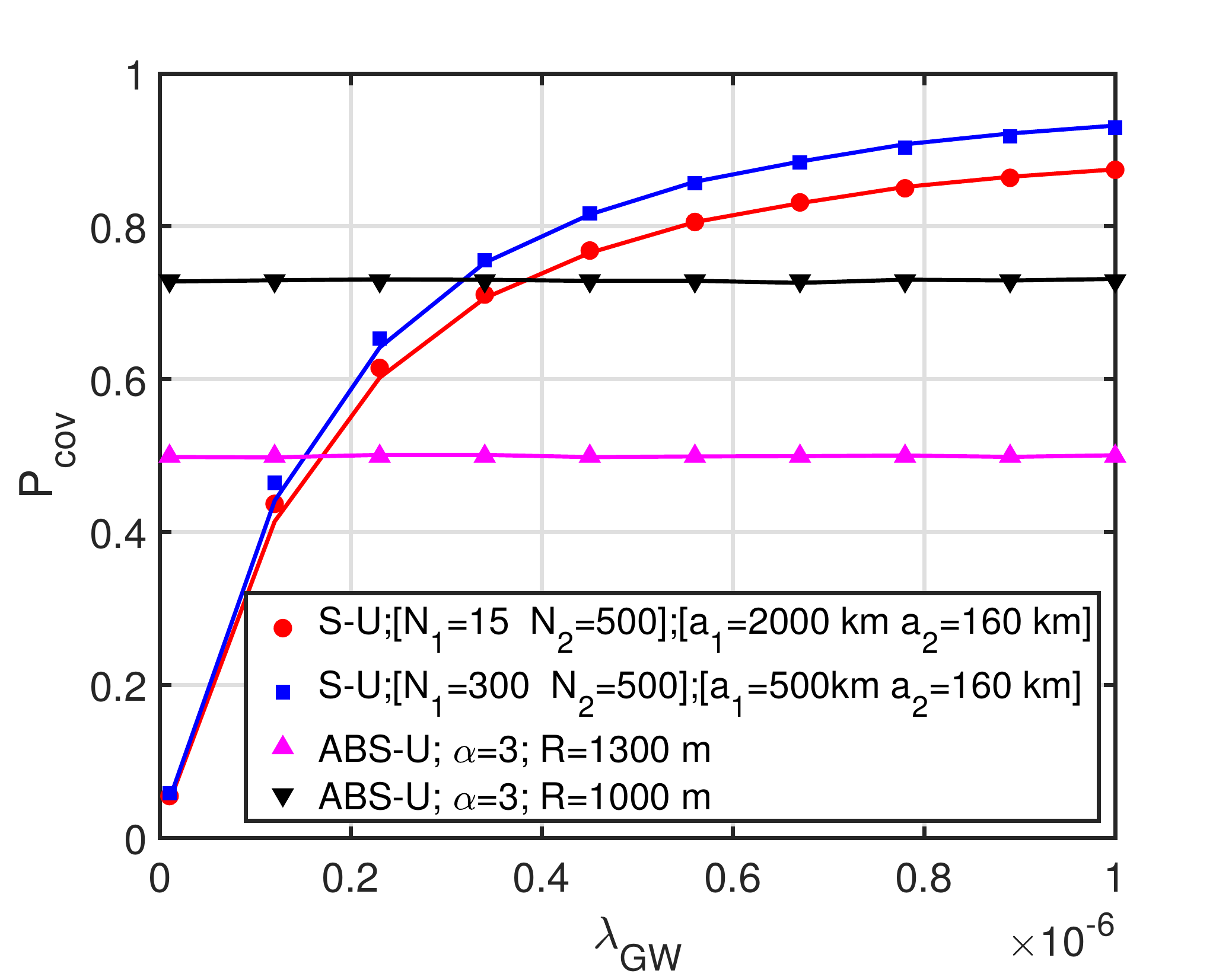}\label{4c}}
\caption{Coverage probability versus density of PPP for GWs, $\lambda_{\rm GW}$}
\label{fig:4}
\end{figure*}
\begin{figure*}[!t] 
\centering
\subfigure[]{\includegraphics[width=0.3\textwidth]{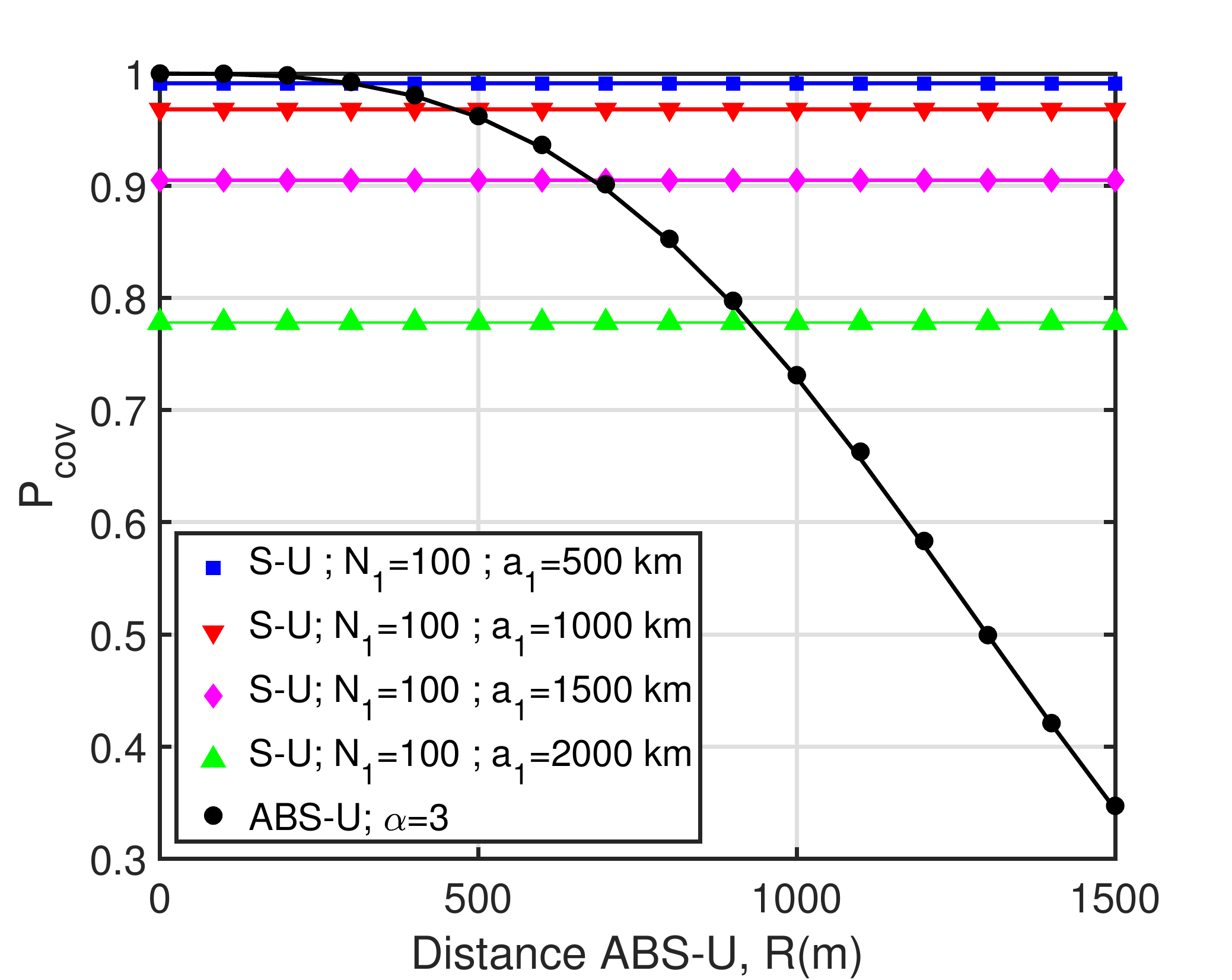}\label{5a}}
\subfigure[]{ \includegraphics[width=0.3\textwidth]{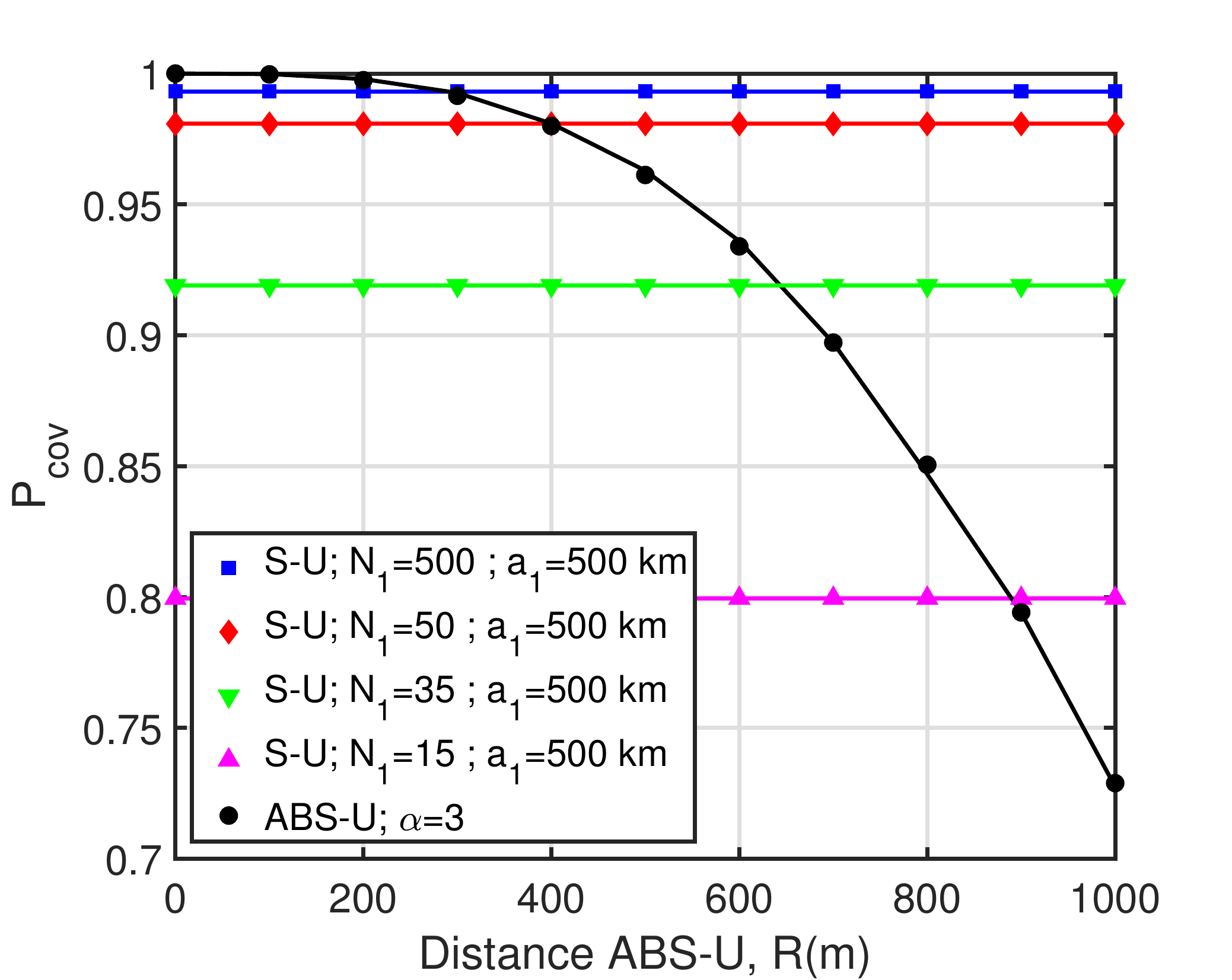}\label{5b}}
\subfigure[]{\includegraphics[width=0.3\textwidth]{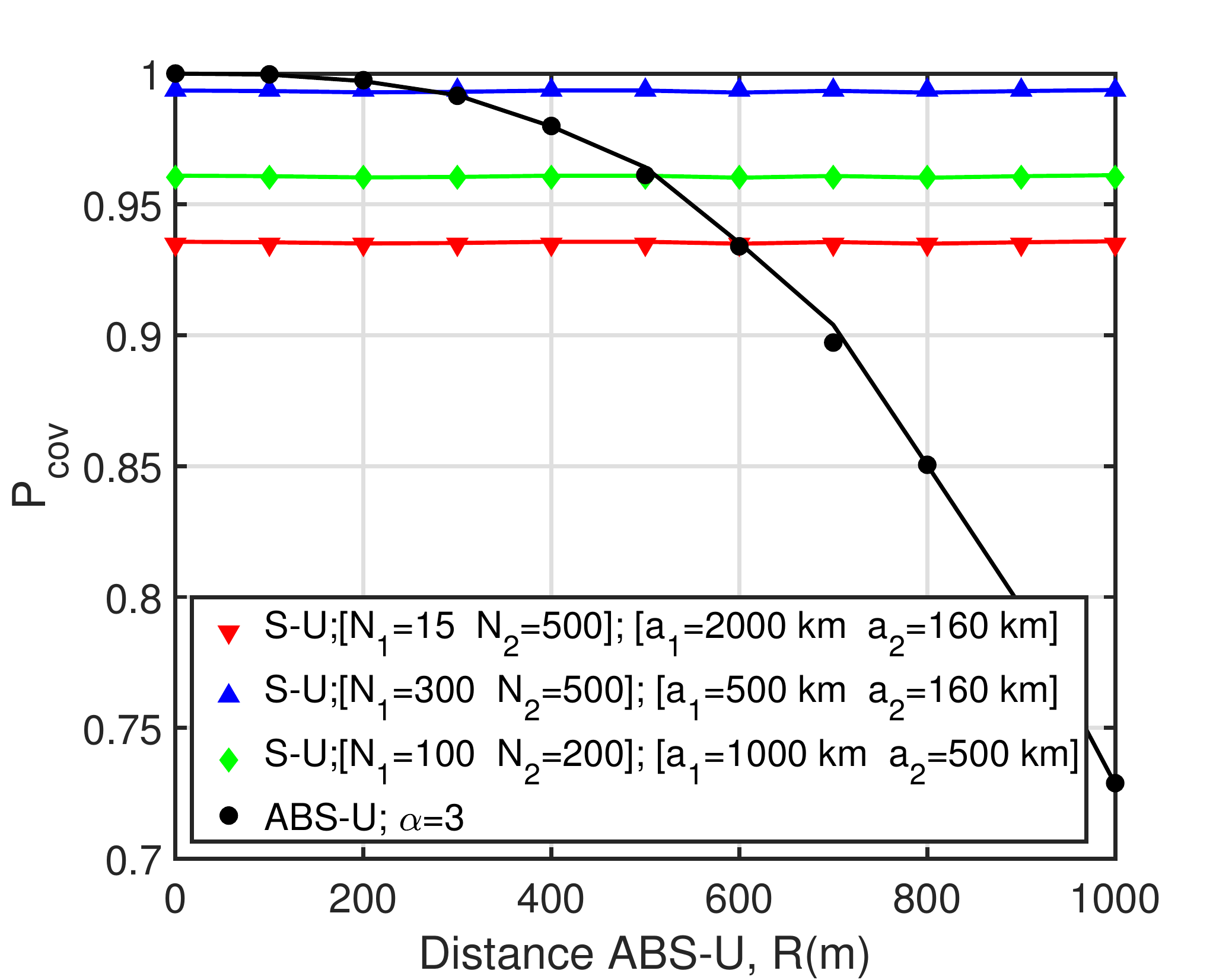}\label{5c}}
\caption{Coverage probability versus distance ABS-U, $R$}
\label{fig:5}
\end{figure*}
In this section, we verify the derived expressions using Monte-Carlo simulations. In addition, we study the influence of various system parameters on the performance of the considered system and on the performance gain in rural and remote areas. The system parameters used in the simulations are summarized in Table~\ref{tab:System_parameters}. In addition, we assume that $\gamma_g=\gamma_u$ (both are denoted as $\gamma_{th}$ in the figures) and $\rho_s=\rho_g=\rho_a$. In all the figures, markers represent the derived analytical results while the solid lines represent the Monte-Carlo simulations.

In Fig.~\ref{3a}, we show that the S-U coverage probability is limited by the GW-U coverage at low values of $\gamma_{th}$, while at higher values it is limited by the $S-U$ coverage. In Fig.~\ref{3b}, we plot the coverage probability for $n=1$ and fixed number of satellites ans study the effect of increasing the altitude. The results show improvement in the coverage probability at lower values of the altitude $a_1$. In Fig.~\ref{3c}, we fix the altitude and vary the number of satellites for the scenario of $n=1$. The results show that the improvement in the coverage probability saturates as we increase the number of satellites beyond a specific value.

In Fig.~\ref{4a}, plot the coverage probability for different values of $\lambda_{\rm GW}$ and $a_1$ while fixing the number of satellites. The results show that the density of gateways required to outperform the ABS coverage reduces as we reduce the altitude of the satellites. Similarly, in Fig.~\ref{4b}, we observe that the required density of gateways reduces as we increase the number of satellites, for a fixed altitude. In Fig.~\ref{4c}, we plot the coverage probability for the scenario of $n=2$ with different values of $a_1$, $a_2$, $N_1$, and $N_2$.

In Fig.~\ref{fig:5}, we show how the satellite communication system improves the coverage in a remote location as the distance to the nearest ABS increases.


\section{Conclusion}
In this paper, we proposed a stochastic geometry-based model for the LEO satellite locations in order to study and analyze the performance of satellite communication systems. Assuming randomly located satellites on spherical surfaces, we derived the coverage probability for a setup where satellite gateways are distributed according to a PPP on the ground. We have verified all the derived expressions using Monte-Carlo simulations and ensured perfect fit. Finally, we have studied the effect of the altitudes of the satellites, their numbers, and the density of the gateways on the performance of the system. The provided framework can be extended in many directions. For instance, the provided framework can be extended to capture more general setups, typically referred to as integrated satellite-aerial-terrestrial networks~\cite{8403964}. 

\begin{table}[!t]
   \centering
    \caption{ System Parameters}
    \label{tab:System_parameters}
    \resizebox{0.4\textwidth}{!}{
\renewcommand{\arraystretch}{0.8}
\begin{tabular}{c|c|c}
\hline Notation & PARAMETER & VALUE \\
\hline 
\hline$f_c$ &  S-GW Link frequency band & $20\mathrm{GHz}(\mathrm{Ka})$ \\
\hline $\rho_s$ & Satellite transmit power & $15 \mathrm{dBW} $\\
\hline $\sigma_g^2$; $\sigma_u^2$ & Noise power at the GW and the user & $3.6\times 10^{-12}$; $10^{-8}$\\
\hline $G_R^2$ & User antenna gain & $41.7 \mathrm{dBi}$ \\
\hline $s$ & Average rain attenuation & $-3.125 \mathrm{dB}$ \\
\hline $\lambda_{\rm GW}$ &Density of GWs & $10^{-5}$ \\
\hline $\alpha$ & path loss exponent & $3$\\
\hline $\mathcal{S}\mathcal{R}\left(\Omega, b_{0}, m\right)$ & SR fading & $\mathcal{S}\mathcal{R}\left(1.29, 0.158, 19.4\right)$\\
\hline
\hline
\end{tabular}}
\end{table}

\appendix
\subsection{Proof of Theorem~\ref{thm:2}}\label{app:1}
\small{\begin{align*}
P_{cov}^{S-GW}& =\operatorname{P}\left(\frac{\rho_r^g}{\sigma_g^2} \geq \gamma_{g} \right)\\
&=\operatorname{P}\left(\frac{\rho_s\left|A\right|^{2} W_{s}^{2}}{\sigma_g^2} \geq \gamma_{g}\right) \\
&=\operatorname{P}\left(\rho_s\Bigg |s\frac{\lambda G_R \xi e^{j \phi}}{4\pi D \sigma_g}\Bigg |^2 W_{s}^2 \geq \gamma_{g}\right)\\ 
&= \operatorname{P}\left(\frac{W_{s}^2}{D^2}\geq c  \right)   \qquad \text{ let  $c=\frac{16 \pi^2 \gamma_{g} \sigma_g^2}{\rho_s| \lambda G_R s e^{j \phi} \xi|^2}$}\\
&= \operatorname{P}\left(\frac{X}{Y} \geq c\right) \qquad \text{ with $W_{s}^2=X$  and  $D^2=Y$}\\
&=\int_{0}^{\infty} \Bar{F}_{X}(cy)f_{Y}(y)dy\\
&=\int_{0}^{\infty} \left(1-{F}_{X}(cy)\right)f_{Y}(y)dy\\
&=\int_{0}^{\infty}f_{Y}(y)dy-\int_{0}^{\infty}\left(\frac{2 b_{0} m}{2 b_{0} m+\Omega}\right)^{m} \sum_{z=0}^{\infty} \frac{(m)_{z}}{z!\Gamma(z+1)}\\
& \times \left(\frac{\Omega}{2 b_{0}m+\Omega}\right)^{z} \gamma\left(z+1, \frac{1}{2 b_{0}} cy\right)f_{Y}(y)dy\\  
&=\int_{0}^{\infty}f_{Y}(y)dy-\left(\frac{2 b_{0} m}{2 b_{0} m+\Omega}\right)^{m} \sum_{z=0}^{\infty} \frac{(m)_{z}}{z ! \Gamma(z+1)}\\ &\times\left(\frac{\Omega}{2 b_{0}
m+\Omega}\right)^{z} \int_{0}^{\infty} \gamma\left(z+1, \frac{1}{2 b_{0}} cy\right)\frac{1}{2 \sqrt{y}}f_{D}(\sqrt{y})dy.
\end{align*}}
\balance
This concludes the proof.

\bibliographystyle{IEEEtran}
\bibliography{Draft_v0.3.bbl}

\end{document}